\documentclass[%
 reprint,
superscriptaddress,
nofootinbib,
 amsmath,amssymb,
 aps,
pra,
]{revtex4-1}
\usepackage{refcount}
\usepackage{graphicx}
\usepackage{dcolumn}
\usepackage{bm}
\usepackage{color}
\usepackage{mathrsfs}
\usepackage[table]{xcolor}

\begin{document}

\preprint{APS/123-QED}

\title{Generating GHZ states with squeezing and post-selection}

\author{Byron Alexander}
 \email{alxbyr001@myuct.ac.za}
 \affiliation{%
 Department of Physics,\\
 Stellenbosch University, Stellenbosch Central 7600, Stellenbosch, South Africa 
}
\author{Hermann Uys}%
 \email{hermann.uys@gmail.com}
\affiliation{%
 Department of Physics,\\
 Stellenbosch University, Stellenbosch Central 7600, Stellenbosch, South Africa 
}
\affiliation{%
Council for Scientific and Industrial Research,\\
National Laser Centre, Brummeria, Pretoria, 0184, South Africa 
}
\author{John J. Bollinger}
 \email{john.bollinger@nist.gov}
\affiliation{
 National Institute of Standards and Technology, Boulder, Colorado, USA\\}%

\date{\today}

\begin{abstract}
Many quantum state preparation methods rely on a combination of dissipative quantum state initialization, followed by unitary evolution to a desired target state. Here we demonstrate the usefulness of quantum measurement as an additional tool for quantum state preparation. Starting from a pure separable multipartite state, a control sequence, which includes rotation, spin squeezing via one-axis twisting, quantum measurement and post-selection, generates a highly entangled multipartite state, which we refer to as \textit{Projected Squeezed} states (or $PS$ states).  Through an optimization method, we then identify parameters required to maximize the overlap fidelity of the $PS$ states with the maximally entangled Greenberger-Horne-Zeilinger states (or $GHZ$ states).  The method leads to an appreciable decrease in state preparation time of $GHZ$ states when compared to preparation through unitary evolution with one-axis twisting only.
\end{abstract}

\pacs{Valid PACS appear here}
\maketitle


\section{\label{sec:level1}Introduction}

Emergent technologies, such as quantum computing, quantum
communication, and quantum sensing, rely principally
on quantum phenomena such as superposition and entanglement for their unique capabilities. These phenomena allow quantum computational devices to overcome
limits set by their classical counterparts in computational
speed of complex algorithms. Furthermore, quantum sensors \cite{degen2017quantum}, devices which utilize quantum correlations to improve sensitivity of measurement by suppressing phase noise in multiparticle interferometry \cite{giovannetti2011advances,hosten2016measurement,gross2012spin}, demonstrate the potential of quantum-enhanced technology. Examples include enhanced performance in atomic clocks  \cite{kessler2014heisenberg,komar2014quantum}, magnetic field detection \cite{ruster2017entanglementmagnetometry} and precision of frequency measurements \cite{huelga1997improvement, ciurana2017entanglement}.

To this end, it becomes
paramount to develop well-defined and efficient protocols to produce
and further exercise control over states of quantum
bits that exhibit desired quantum mechanical traits. Our investigation focuses on establishing a protocol that uses quantum control operations
combined with measurement and post-selection to produce highly entangled metrologically relevant states.  We will refer to these states as  \textit{Projected
Squeezed} states (see \cite{walter2016multi,toth2014quantum, coffman2000distributed,meyer2002global} for relevant discussions on measures of multipartite entanglement). We further study optimization of the control parameters that produce maximal overlap
of the projected squeezed state with the well-known
\textit{Greenberger-Horne-Zeilinger} state (commonly referred to
as the maximally entangled state or $GHZ$ state, see \cite{greenberger1989going, walter2016multi}). For a multipartite system consisting of $N$-qubits, the $GHZ$ state reads as follows
\begin{align}
|GHZ\rangle := \frac{|0\rangle^{\otimes N} + |1\rangle^{\otimes N} }{\sqrt{2}}.
\end{align}
Due to their high level of entanglement, the $GHZ$ states
are of importance in various applications such as metrology \cite{toth2014quantum}, quantum teleportation \cite{gorbachev2000quantum}, quantum computing \cite{gottesman1999demonstrating} and quantum secret sharing \cite{hillery1999quantum}. There are numerous
proposed schemes for producing $GHZ$-states particularly
in the context of cavity quantum electrodynamics \cite{guerra2005realization,zhang2014generation,izadyari2016creation,zheng2001one,chen2012one,zhang2017generation}. Some of the most successful implementations have been in trapped-ion systems, where $14$-ion $GHZ$ states \cite{monz201114} and more complex entangled states of up to $20$ ions \cite{friis2018observation} have been observed. Recently 20-qubit $GHZ$ states have been generated through unitary evolution with Rydberg atom qubits \cite{omran2019generation} and superconducting circuit qubits \cite{song2019observation}. Using post-selection in a linear optical system, $GHZ$
states of $10$ photons have been reported \cite{wang2016experimental}. Closely related to the photon $GHZ$ states are the
so-called $NOON$ states, which also exhibit an improvement
on the standard quantum limit with regard to phase error measurements \cite{dowling2008quantum}. A number of proposed schemes
for producing $NOON$ states exist \cite{kapale2007bootstrapping,mitchell2005metrology,kok2002creation,pryde2003creation}.

Our approach expands the typical suite of quantum state preparation tools, which  relies on dissipative state initialization followed by unitary evolution, to include quantum measurement. The particular example illustrates that non-trivial speed-up can be achieved as compared to state preparation with unitary evolution only. This aspect may be of interest to beat decoherence time-scales in appropriate scenarios. Measurement-based state preparation has been discussed and demonstrated for spin-squeezed states (\cite{cox2016deterministic,schleier2010states}). Only limited investigations have been  carried out for more general state preparation protocols (for examples see \cite{nielsen2003quantum,yan2015quantum}).

The setup we have
in mind is an ensemble of two-level systems, with eigenstates represented in the collective pseudo-spin basis (also known as the \textit{Dicke state basis} \cite{dicke1954coherence}). The projected squeezed state is
produced through a sequence of control operations including initialization,
rotation, spin squeezing \cite{kitagawa1993squeezed}, quantum measurement
and post-selection. Experimentally, the main technical
challenge is carrying out a projective measurement of the
collective spin projection quantum number (as opposed
to a measurement in the single particle basis), as all other
aspects are well established. 

The one-axis twisting spin-squeezing operator (also known as the \textit{Kitagawa Shearing Gate}), which was introduced in a seminal paper \cite{kitagawa1993squeezed}, is described by the following unitary transformation
\begin{align}
\hat{U}_{Sq}(t) = \text{exp}(-i \chi t \hat{J}^2_{z}).\label{Sq}
\end{align}
One-axis twisting has been realized with trapped ions \cite{uys2012toward}, neutral atoms \cite{hu2017vacuum,mussel2014scalable} and superconducting circuits \cite{song2019observation}.
Here $\chi$ quantifies the strength of the squeezing interaction, and
\begin{align}
\hat{J_k} := \sum_{i}^{N}\frac{1}{2}\sigma_{i}^{k},
\end{align}
where $k=x,y,z$ and $\sigma^k_i$ is the $k$-component of the usual Pauli spin operator for the $i$'th two-level system in an ensemble of $N$ systems. This definition preserves the spin commutation relation $[\hat{J_x}, \hat{J_y}] = 2i \epsilon_{xyz}\hat{J_z}$ for the pseudo-spin $\hat{J^2}=\hat{J^2_x}+\hat{J^2_y}+\hat{J^2_z}$. In what follows, we will restrict ourselves to a subspace of the full pseudo-spin Hilbert space, namely the fully symmetric (Dicke) eigenstates for which:
\begin{align*}
\hat{J_z}\bigg|\frac{N}{2}, \frac{N}{2}-m \bigg\rangle = \bigg(\frac{N}{2}-m \bigg)\bigg|\frac{N}{2}, \frac{N}{2} - m \bigg \rangle,
\end{align*}
with $m = 0,1,2,...N$ and 
\begin{align*}
\hat{J^2}\bigg|\frac{N}{2}, \frac{N}{2}-m \bigg\rangle = \bigg(\frac{N}{2}\bigg)\bigg(\frac{N}{2}+1\bigg)\bigg|\frac{N}{2}, \frac{N}{2}-m \bigg\rangle.
\end{align*}
A method for constructing the Dicke states  from the single spin basis is discussed in \cite{uys2012toward}.



\section{\label{sec:level1}Method}
We now describe the steps in the state preparation protocol. As an initial state we choose the pure, separable state 
\begin{align}
| \psi(0) \rangle  = \bigg|\frac{N}{2}, \frac{N}{2} \bigg\rangle.
\end{align}
The protocol, in sequence, consists of the following operations: 
\\
\\
\textit{Step 1} - An initial rotation by $\frac{\pi}{2}$ about the $x$-axis to form what is known as the \textit{coherent spin} state (or $CS$ state)
\begin{align}
| CS \rangle = \frac{1}{2^{N/2}}\sum_{M = -\frac{N}{2}}^{\frac{N}{2}}\binom{N}{\frac{N}{2} + M}^{1/2}\bigg|\frac{N}{2},M \bigg\rangle.
\end{align}
We can visually represent any state, $| \psi \rangle$, on the Bloch-sphere by considering the modulus squared of the projection of that state  onto a rotated coherent spin state, $H=|\langle\psi|\text{exp}(-i \phi \hat{J}_{z}) \text{exp}(-i \theta \hat{J}_x) | CS\rangle|^2$, where $\theta$ and $\phi$ are respectively the polar and azimuthal angles. These are commonly referred to as \textit{Husimi plots} \cite{lee1984q}.  This projection, when $|\psi\rangle=|CS\rangle$, is shown on a unit sphere in Fig.~1(a). It shows that the rms width of $H$ is uniform in all directions for this case. \\

\noindent
\textit{Step 2} - The coherent spin state then undergoes \textit{squeezing} by the unitary transformation Eq.~(\ref{Sq}), where the magnitude of squeezing is controlled by choices of the squeezing parameter $\chi t$  \cite{uys2012toward}. This is shown in Fig.~1(b)-(c) for different values of $\chi t$.
As we can see, when acting on a coherent spin state, the squeezing operator reduces the spin uncertainty along one spin axis at the expense of increasing the uncertainty along an orthogonal spin axis. The reduction in uncertainty occurs symmetrically about an axis tilted slightly with respect to the $x$-axis as opposed to the $x$-axis itself. 
\begin{figure*}
\hspace*{-1.72cm}
  \centering
  \begin{tabular}{ccc}
    \includegraphics[width=65.5mm,scale=0.9]{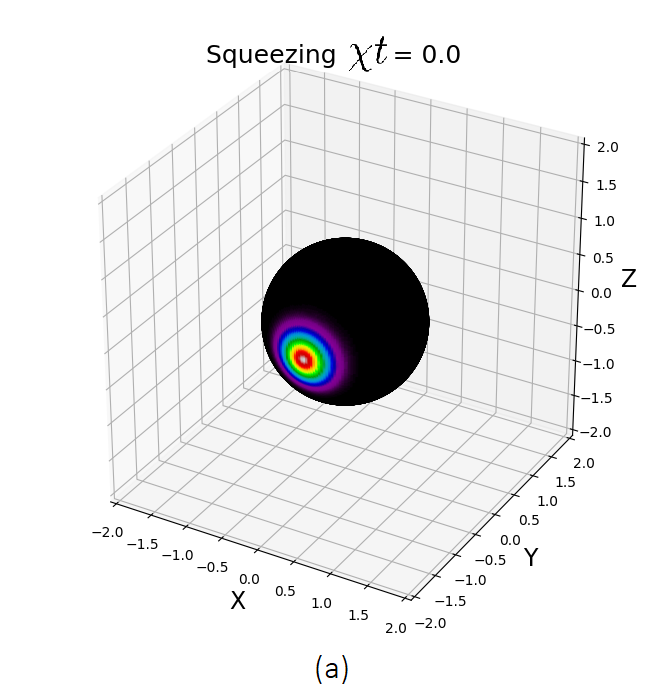}& \includegraphics[width=67.5mm,scale=0.99]{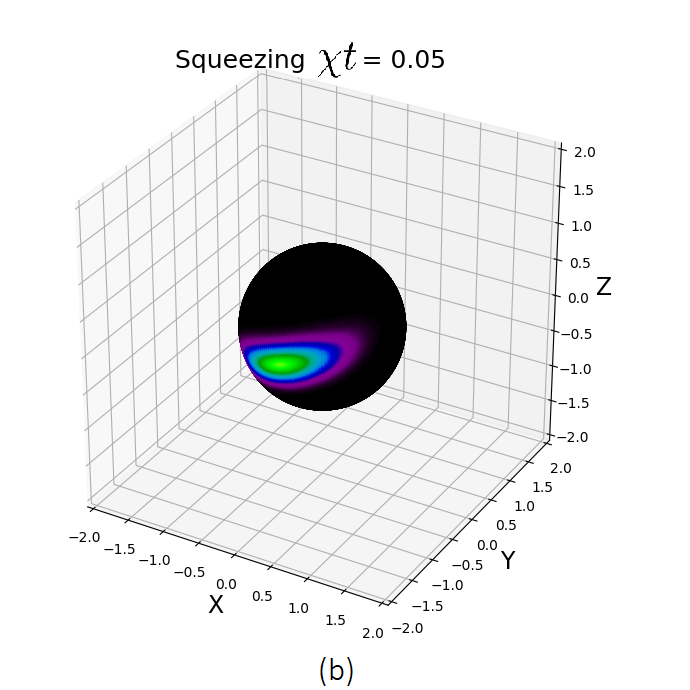}& \includegraphics[width=67.5mm,scale=0.99]{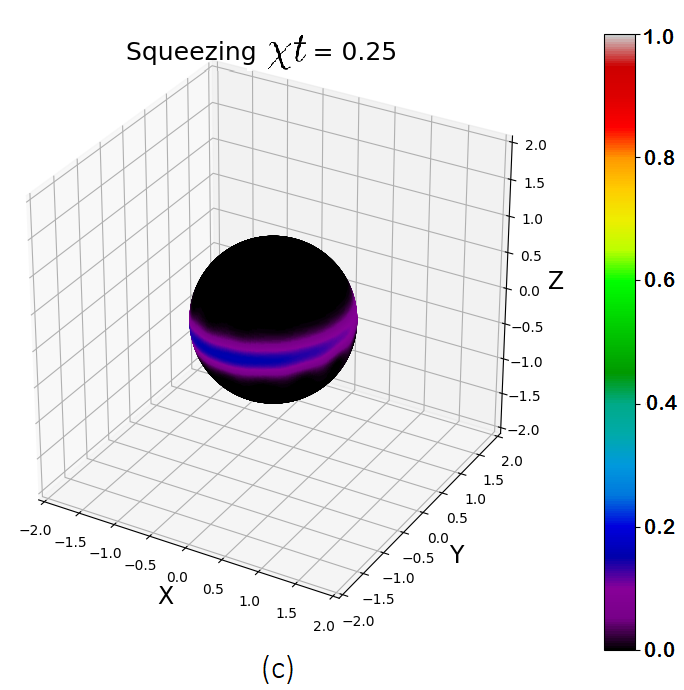}
  \end{tabular}
  \caption{Overlap fidelity of squeezed state with rotated $CS$ state projected onto a sphere; for varied squeezing ($N=50$).}
  \label{A}
\end{figure*}
A choice of the squeezing parameter of approximately $\chi t=0.25$ starts producing a projection sufficiently flat so as to create a probability ring that wraps around the sphere (as shown in Fig.~\ref{A}(c)). 
\\
\\
\textit{Step 3} - Following the squeezing, we rotate about the $x$-axis until the ring is aligned with the $z$-axis as shown in Fig.~\ref{A4}. 
\begin{figure}[htp]
\hspace*{-0.35cm}
  \centering
  \begin{tabular}{ccc}
   \includegraphics[width=99mm,scale=0.995]{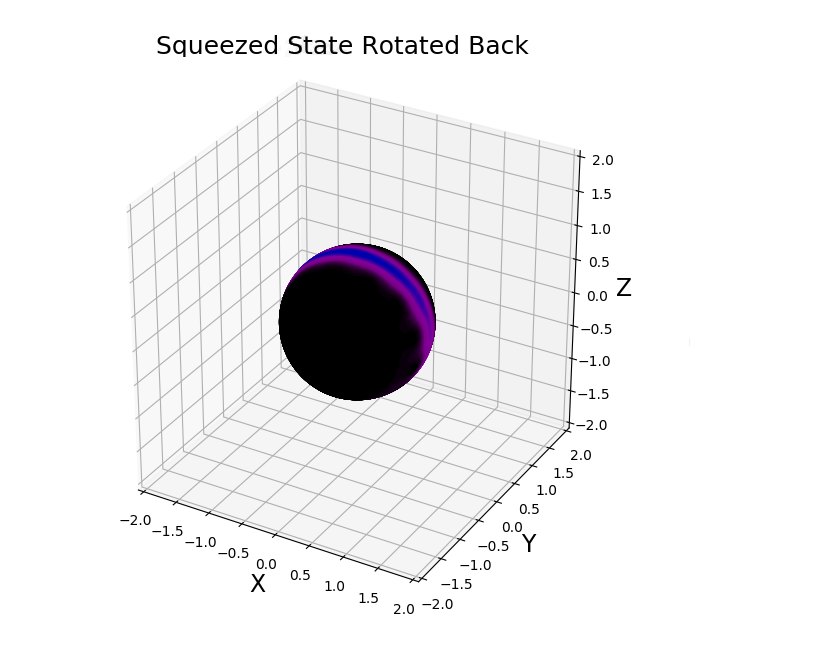}
  \end{tabular}
  \caption{Overlap fidelity with rotated \textit{CS} state, after rotation, squeezing ($\chi t = 0.25$) and rotation back.}
  \label{A4}
\end{figure}
\\
\\
\textit{Step 4} - The appropriate quantum measurement is carried out, and the desired state is post-selected based on the measurement outcome. The Kraus operators that describe our quantum measurement are chosen as follows:
\begin{align}
 A_C := \sum_{M}\sqrt{ \text{Pr}(M|C)}\bigg|\frac{N}{2}, M \bigg\rangle \bigg\langle \frac{N}{2}, M \bigg|, \label{Ac}
\end{align}
with \textit{Gaussian probability distribution}
\begin{align}
\text{Pr}(M|C):= \frac{1}{\sqrt{2 \pi \sigma^2}}\exp \bigg(-\frac{(M-C)^2}{2 \sigma^2} \bigg).\label{Pr}
\end{align}

As required, the operators $A_C$ obey the normalization condition $\int A_{C} A_{C}^{\dagger}dC = I$. Physically, a measurement of $A_{C'}$, with outcome $C^\prime$, will project an initial wave function onto a superposition of states with amplitudes following a Gaussian distribution and centered on $C^\prime$, with width $\sigma$.

Since the set of allowed measurement outcomes $\{C\}_{C \in \mathbb{R}}$ is continuous, the resultant quantum state after measurement is thus given by 
\begin{equation}
\rho \mapsto \tilde{\rho}_{\text{final}} = \frac{A_C \rho A_{C}^{\dagger} dC}{\text{Tr}[A_C \rho A_{C}^{\dagger}dC]},\label{rotilde}
\end{equation} 
where $\rho$ denotes the density matrix that describes the state of our system, and $\text{Tr}[A_C \rho A_C^{\dagger}dC]$ the probability density to observe a measurement outcome in the interval $[C,C+dC]$ (see \cite{jacobs2014quantum}).
For computational purposes we have to discretize the distribution $\text{Pr}(M|C)$ by binning the $C$-axis and integrating over each bin to obtain probabilities instead of probability densities, thus allowing us to model the measurement statistics numerically. 
   
\begin{figure}[h!]
  \centering
  \hspace*{-0.7cm}
  \begin{tabular}{ccc}
    \includegraphics[width=87mm,scale=0.90]{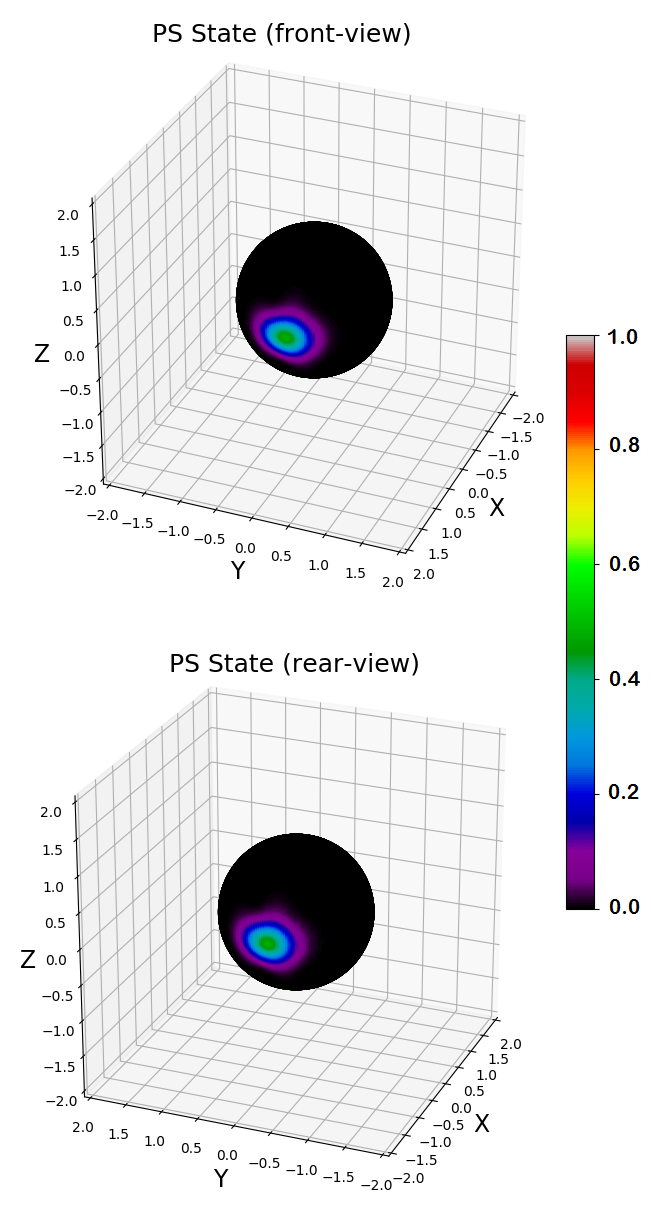}& 
   \end{tabular}
  \caption{Overlap fidelity of \textit{PS} state and rotated \textit{CS} state (\textit{N}=50).}
  \label{B}
\end{figure}

To generate the desired state, the quantum measurement defined by Eq. (\ref{Ac}) is executed, and only outcomes with $C \approx 0$ are post-selected. This produces what we refer to as a \textit{projected squeezed} state, henceforth denoted $|PS\rangle$. The resultant state after measurement, as shown in Fig.~\ref{B}, consists of two probability lobes concentrated on opposing sides of the Bloch-sphere. Here we used $N = 50$, $\chi t = 0.4$ and measurement operator variance $\sigma^2 = 22$ for optimization reasons which will be discussed shortly.
\\
\\
\textit{Step 5} - Finally, we generate a state which closely resembles the $GHZ$ state by executing a rotation by $\frac{\pi}{2}$ about the $y$-axis. Then, the resemblance to the $GHZ$ state is quantified by computing the measure of `closeness' of two pure quantum states: $\mathcal{F} = |\langle PS | GHZ \rangle |^2$. $\mathcal{F}$ is known as the \textit{overlap fidelity}.

\begin{figure}[tp]
\hspace*{-0.35cm}
  \centering
  \begin{tabular}{ccc}
   \includegraphics[width=92mm,scale=0.95]{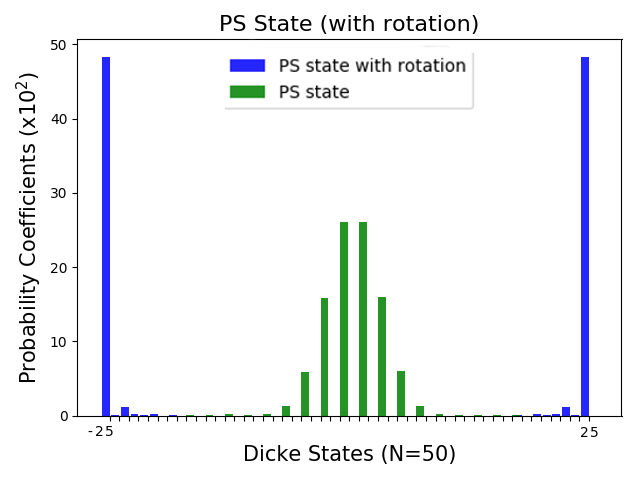}
  \end{tabular}
  \caption{Probability coefficients of \textit{PS} state in Dicke State basis.}
  \label{D}
\end{figure}

For completeness, we plot in Fig.~\ref{D}, the modulus squared of the probability distribution of the $PS$ state in the Dicke basis as generated in step 4 of the protocol, and after the rotation about the $y$-axis in step 5. It shows that any imperfect overlap is due to the unintended occupation of close-lying states other than $|\frac{N}{2}, \pm \frac{N}{2} \rangle$ with small probability amplitudes. 

\section{\label{sec:level2}Optimization}
A numerical optimization method (random walk MCMC - Markov chain Monte Carlo type regime, see \cite{gilks1995markov,sherlock2010random}) is now employed to find parameters of $\sigma^2$ and $\chi t$ that maximize the overlap fidelity with the $GHZ$ state.

Given initial values of $\sigma^2$ (the variance used in defining the measurement operators) and $\chi t$ (the squeezing parameter), we define an initial vector $(\sigma^2_0, {\chi t}_0)$. The numerical algorithm stochastically traverses the parameter space in steps defined by the vector
\begin{align*}
(\sigma^2_{step_1}, \chi t_{step_1}) := (\sigma^2_0, {\chi t}_0) + (d \sigma^2, d \chi t).
\end{align*} The increments $d \sigma^2$ and $d \chi t$ are random variables in that they are respectively chosen from Gaussian probability distributions centered at zero (with the variance of these Gaussian distributions appropriately chosen to minimize the time of computation). For step $n$, the overlap fidelity is computed for parameter values $( \sigma^2_{step_n}, \chi t_{step_n})$. If the fidelity is increased, the new vector is retained, otherwise we reject the step and retain the previous vector $(\sigma^2_{step_{n-1}}, \chi t_{step_{n-1}})$. Subsequently, we compute a new step and again compare this step to the previous step. This process is continued until we identify parameters which produce an overlap fidelity value greater than or equal to a fixed threshold value. With $N=50$, this optimization leads to maximum of $\mathcal{F}=0.97$ for the parameters $\chi t \approx 0.4$ and $\sigma^2 = 22$.

\section{\label{sec:level3}Analysis and Efficiency}
\begin{figure}[htp]
\hspace*{-0.25cm}
  \centering
  \begin{tabular}{ccc}
   \includegraphics[width=97mm,scale=0.91]{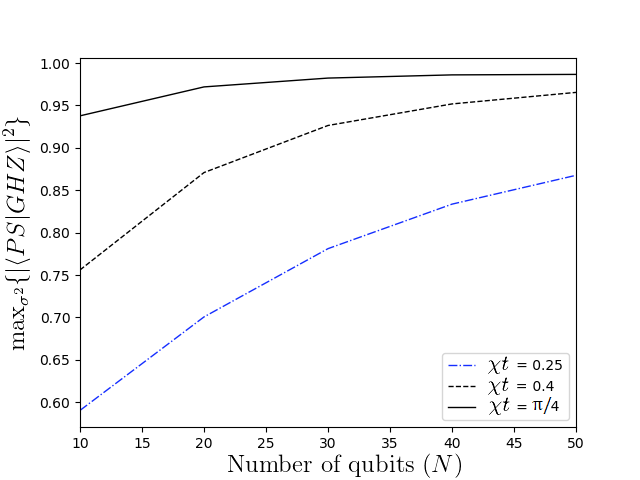}
  \end{tabular}
  \caption{Maximal $GHZ$ overlap fidelity for varied squeezing times.}
  \label{G9}
\end{figure}

To map out cross-sections of the optimization landscape, we fixed individual parameters (after they have been optimized) while allowing the others to vary. This firstly shows that the maximum fidelity monotonically increases with increasing particle number $N$, as illustrated in Fig.~\ref{G9}, for different values of $\chi t$.

\begin{figure}[htp]
\hspace*{-0.8cm}
  \centering
  \begin{tabular}{ccc}
   \includegraphics[width=99mm,scale=0.91]{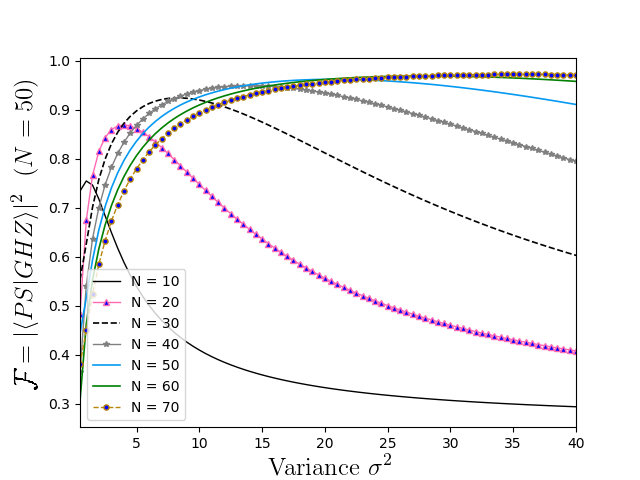}
  \end{tabular}
  \caption{Overlap fidelity of $PS$ and $GHZ$ states as a function of $\sigma^2$; for varied $N$ $(\chi t = 0.4)$.}
  \label{H}
\end{figure}

\begin{figure}[htp]
\hspace*{-0.63cm}
  \centering
  \begin{tabular}{ccc}
   \includegraphics[width=99mm,scale=0.91]{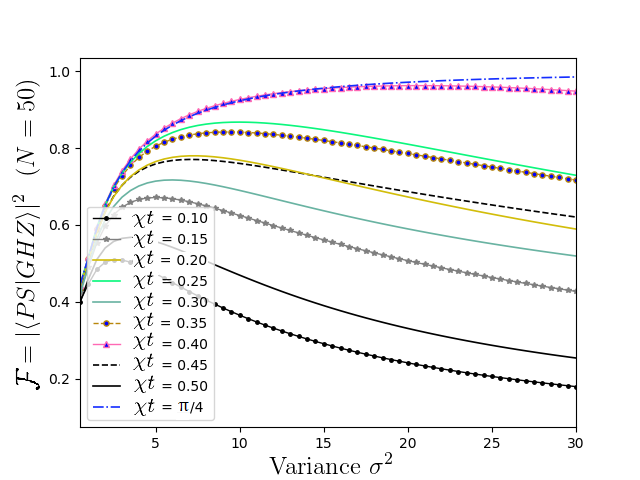}
  \end{tabular}
  \caption{Overlap fidelity of $PS$ and $GHZ$ states as a function of $\sigma^2$; for varied squeezing times $(N=50)$.}
  \label{E1}
\end{figure}

In Fig.~\ref{H}, $\chi t$ is fixed at $0.4$, and the fidelity is plotted as a function of $\sigma^2$ for different particle numbers. It confirms that the maximum fidelity increases with $N$, and shows that at larger particle number, the protocol is much less sensitive to variations in $\sigma$, producing high fidelity over wider regions of the variance.
Figure \ref{E1} shows the fidelity as a function of variance for fixed $N=50$, and for different values of the squeezing $\chi t$. Around $\sigma^2=10$, there are local maxima in $\mathcal{F}$ at squeezing $\chi t \approx 0.25, 0.4$ and $\pi/4$.

It is important to note that an exact $GHZ$ state can be produced (with $\mathcal{F} = 1$) by using only the squeezing interaction with $\chi t = \frac{\pi}{2}$ and a rotation\footnote{In principle the $PS$ state protocol could be employed to produce a state with $\mathcal{F} \to 1$ as $\sigma^2 \to \infty$ (for $\chi t = \frac{\pi}{2}$). This is clear since as the variance $\sigma^2 \to \infty$, the Gaussian distribution (\ref{Pr}), which defines our Kraus measurement operators, tends to a uniform probability distribution and therefore the measurement operator, for $C'=0$, approaches the identity.}. We emphasize that given $\chi t \approx 0.25$ (or $0.4$), our measurement-based protocol produces highly entangled $GHZ$ type states about a factor $6$ (or respectively a factor $4$) faster than the coherent protocol with $\chi t = \frac{\pi}{2}$. As such, this measurement-based protocol may be preferable if a relevant decoherence timescale is close to $\chi t= \frac{\pi}{2}$.

Over and above high overlap fidelity an important consideration is the efficiency with which the $PS$ state is produced.  
We will characterize a state preparation protocol as \textit{efficient} if it requires low squeezing parameter $\chi t$ (hence less time required for squeezing), produces high  overlap fidelity $GHZ$ and, given the inherent stochastic nature of the process, has a high measurement outcome probability. 

\begin{figure}[htp]
\hspace*{-0.3cm}
  \centering
  \begin{tabular}{ccc}
   \includegraphics[width=94mm,scale=0.8]{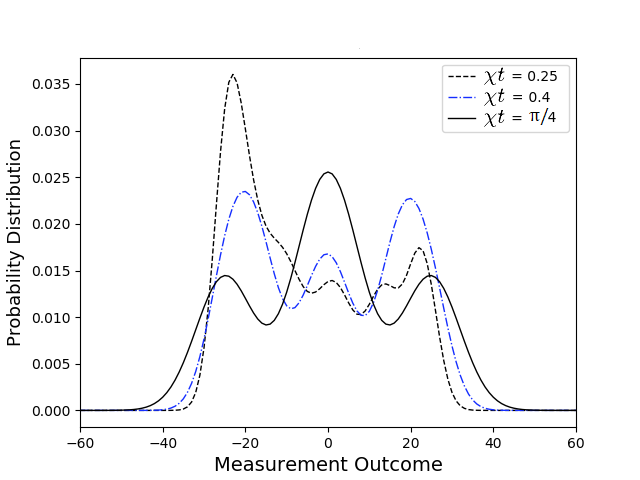}
  \end{tabular}
  \caption{Probability distribution of measurement outcomes  for varied squeezing times ($N = 50$).}
  \label{G1}
\end{figure}

\begin{figure}[htp]
\hspace*{0.0cm}
\vspace*{-0cm}
  \centering
  \begin{tabular}{ccc}
   \includegraphics[width=82mm,scale=0.99]{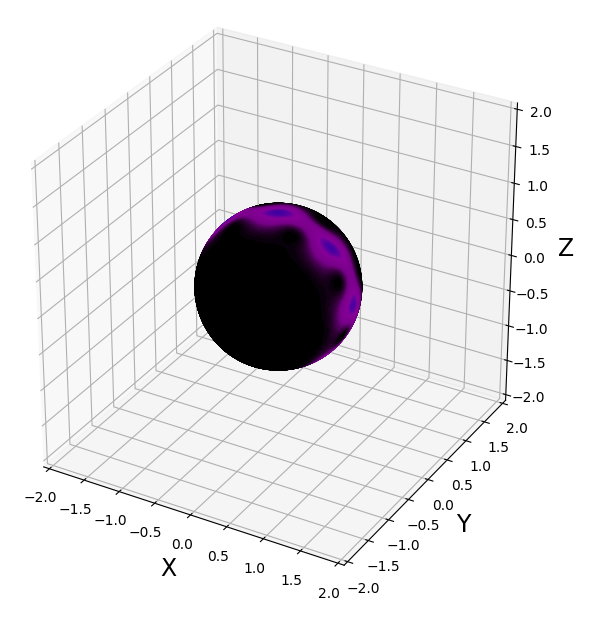}
  \end{tabular}
  \caption{Pre-measurement squeezed state ($\chi t \approx 0.4$) rotated back about $x$-axis.}
  \label{G3}
\end{figure}

\begin{figure}[htp]
\hspace*{0.0cm}
  \centering
  \begin{tabular}{ccc}
  \vspace{-1cm}
   \includegraphics[width=82mm,scale=0.99]{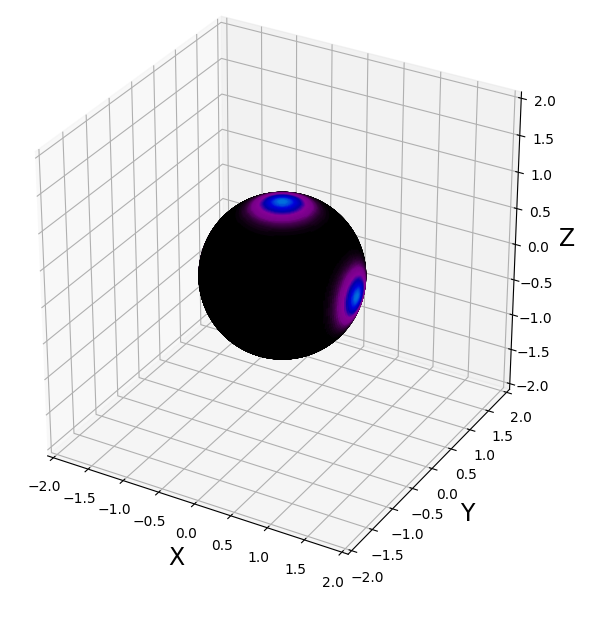}
  \end{tabular}
  \caption{Pre-measurement squeezed state ($\chi t \approx \pi/4$) rotated back about $x$-axis.}
  \label{G10}
\end{figure}

Using the MCMC optimization protocol, we find that maxima in the overlap fidelity between the $PS$ and $GHZ$ states (varying $N$ and $\sigma^2$) occur about squeezing parameters $\approx 0.25,~ 0.4$ and $\frac{\pi}{4}$ ($\chi t = \frac{\pi}{2}$ as stated above, requires no measurement). 
We plot in Fig.~\ref{G1} the probability of obtaining measurement outcomes $\{C\}_{C \in [-60,60]}$ for the aforementioned squeezing times. For comparison, the full pre-measurement Husimi plots are shown in Figs.~\ref{A4},~\ref{G3} and \ref{G10}, respectively.

\begin{figure*} [tpbh]
\hspace*{-1.25cm}
  \centering
  \begin{tabular}{ccc}
    \includegraphics[width=69mm,scale=0.9]{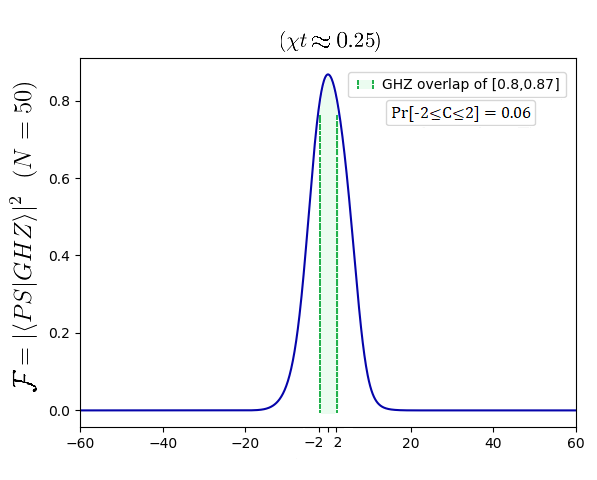}   
    \includegraphics[width=63.5mm,scale=0.9]{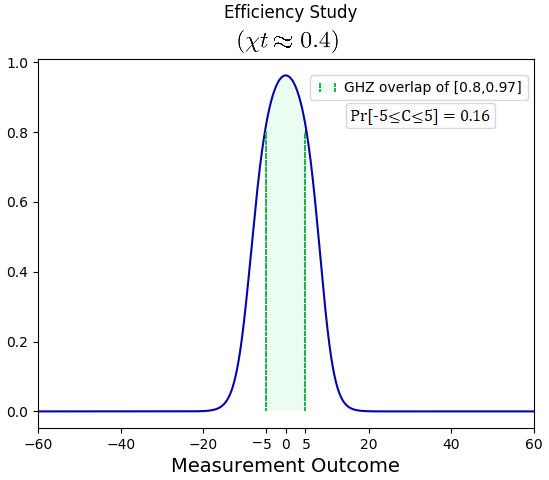}& \includegraphics[width=63.5mm,scale=0.9]{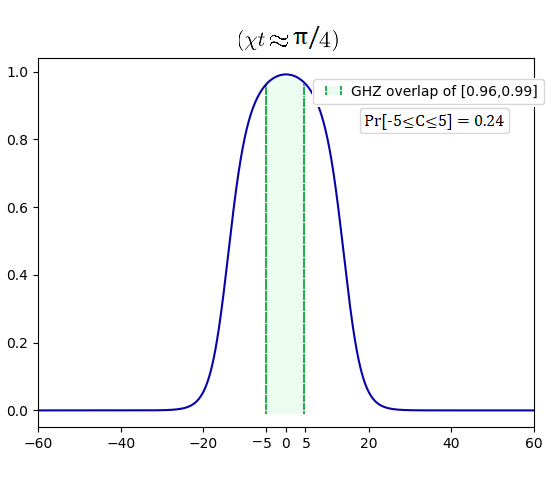}&
  \end{tabular}
  \caption{Selected plots of $\mathcal{F}$ as a function of the measured outcome $C$. The vertical green lines indicate the range of outcomes $C$ that yield $\mathcal{F}$-values falling in the range $[\cdot,\cdot]$ indicated in the text box in each sub-figure. The total likelihood, Pr$[\cdot]$, of observing one of those outcomes is also indicated.}
  \label{TP}
\end{figure*}

There are distinct probability peaks in each of the probability distributions represented in Fig.~\ref{G1}. These peaks are due to the probability lobes seen in the Husimi plots of the rotated squeezed state. The maxima of the central peaks correspond to our desired post-selected outcome $C'=0$.  
Fig.~\ref{TP} plots the overlap fidelity $\mathcal{F}$ for each of the local maxima squeezing parameters and highlights the resultant fidelity (for select measurement outcome intervals). It shows that for $\chi t \approx 0.4$, measurement outcomes in the range $[-5,5]$ have overlap fidelities in the range $[0.80, 0.97]$. The probability of obtaining a measurement result in this range is $0.16$ (approximately $1$ success for every $6$ trials). There is therefore a very reasonable success ratio for projecting on states with at least moderately high overlap with the $GHZ$ state.


 Squeezing parameters $\chi t \approx 0.25,~ 0.4$ and $\pi/4$ respectively yield maximal $PS$ and $GHZ$ state fidelity values, given $C^\prime=0$, of $0.87,~0.97$ and $0.99$. A salient feature of squeezing $\chi t \approx \pi/4$, as compared to $\chi t \approx 0.25\text{ or } 0.4$, is that the desired post-selected measurement outcome $C'=0$ is the most probable outcome (see Fig.~\ref{G1}).

\subsection{Efficiency Results}

In Table I we summarize the efficiency of the protocol by showing the range (codomain) of $\mathcal{F}$ for particular measurement outcome intervals. The analysis gives the resultant efficiency of the protocol given the optimized set of squeezing parameters (with $\sigma^2$ chosen accordingly).

\section{Discussion}
Using the method presented one can create $GHZ$ states with high fidelity in quantum spin systems. In trapped ion and neutral atom systems, state detection conventionally relies on fluorescence scattering from a dipole-allowed closed-cycle transition. This, however, projects the spins in the single-particle basis rather than the Dicke basis, which makes it an unsuitable quantum measurement for our purposes. In trapped-ion systems, one potential method for executing the collective measurement is to do state dependent excitation of the ion motion using the optical dipole force \cite{hosten2016measurement}. The image current induced in the ion trap electrodes is expected to be proportional to the projection quantum number $M$ and not to the individual ion state. This can be used to implement the measurement operator in Eq.~(\ref{Ac}).

Starting with the pure separable state $| \psi \rangle = | \frac{N}{2}, \frac{N}{2} \rangle$, we showed that using a combination of spin squeezing,
quantum measurement and post-selection it is possible
to generate many-particle $GHZ$ states with high fidelity ($\mathcal{F} > 0.99$ given $N =50$).  It is a comparatively efficient method in the sense that, despite its stochastic nature, we produce these highly entangled $PS$ states for squeezing parameter $\chi t$ significantly lower than that required when doing coherent squeezing (\ref{Sq}) only. This may be beneficial for beating decoherence limitations in some experiments.

\begin{figure*}
\hspace*{0.0cm}
\setlength{\tabcolsep}{0.5cm}
    \begin{table*}
        
        \centering
        \hspace*{-0.6cm}
\renewcommand{\arraystretch}{2.0}
        \begin{tabular}{|c|c|c|c|c|c|}
        \hline
    \rowcolor[gray]{0.9}[0.5cm]
        \multicolumn{6}{|c|}{\boldsymbol{$\chi t \approx 0.25~~~~, N = 50~~~~, \sigma^2 = 10.$}}\\
        \hline
        $\mathscr{R}(\mathcal{F})_{C \in [-5,5]}$  & \text{Pr}[$-5 \leq C \leq 5$] & $\mathscr{R}(\mathcal{F})_{C \in [-2,2]}$ & \text{Pr}[$-2 \leq C \leq 2$] & $\mathscr{R}(\mathcal{F})_{C \in [-1,1]}$ & \text{Pr}[$-1 \leq C \leq 1$] \\
        \hline
        [0.46,0.87] & 0.132 & [0.80,0.87] & 0.056 & [0.85,0.87] & 0.029 \\
        \hline
    \rowcolor[gray]{0.9}[0.5cm]
        \multicolumn{6}{|c|}{\boldsymbol{$\chi t \approx 0.4~~~~, N = 50~~~~, \sigma^2 = 22.$}}\\
        \hline
        $\mathscr{R}(\mathcal{F})_{C \in [-5,5]}$  & \text{Pr}[$-5 \leq C \leq 5$] & $\mathscr{R}(\mathcal{F})_{C \in [-2,2]}$ & \text{Pr}[$-2 \leq C \leq 2$] & $\mathscr{R}(\mathcal{F})_{C \in [-1,1]}$ & \text{Pr}[$-1 \leq C \leq 1$] \\
        \hline
        $[0.80,0.97]$ & $0.161$ & $[0.94,0.97]$ & $0.067$ & $[0.956,0.965]$ & $0.035$ \\
        \hline
     \rowcolor[gray]{0.9}[0.5cm]      
        \multicolumn{6}{|c|}{\boldsymbol{$\chi t \approx \pi/4~~~~, N = 50~~~~, \sigma^2 = 49.$}}\\
        \hline
        $\mathscr{R}(\mathcal{F})_{C \in [-10,10]}$  & \text{Pr}[$-10 \leq C \leq 10$] & $\mathscr{R}(\mathcal{F})_{C \in [-5,5]}$ & \text{Pr}[$-5 \leq C \leq 5$] & $\mathscr{R}(\mathcal{F})_{C \in [-2,2]}$ & \text{Pr}[$-2 \leq C \leq 2$] \\
        \hline
        $[0.80,0.99]$ & $0.409$ & $[0.96,0.99]$ & $0.242$ & $[0.987,0.992]$ & $0.104$ \\
        \hline
        \end{tabular}
        \caption{Consider the range (co-domain) $\mathscr{R}(\cdot)$ of the overlap fidelity $\mathcal{F}$ taken over some chosen interval $C \in [\cdot, \cdot]$ of measurement outcomes; and the respective probability Pr$[\cdot]$ of obtaining outcomes in this interval.}. 
        
        \label{tab:EfficiencyA}
    \end{table*}
\end{figure*}

\newpage
\section{Acknowledgements} 
This project was funded by the CSIR and Department of Science and Technology. We thank E. Jordan and R. Lewis-Swan for useful comments and discussions and I. Cirac for bringing this problem to our attention. We are also grateful to Sulona Kandhai, from the University of Cape Town, for fruitful discussions on employing numerical optimization methods.  
\bibliographystyle{ieeetr}
\bibliography{bibfile}

\end{document}